\documentclass{pasa}%
\usepackage{subcaption}
\usepackage{graphicx}
\usepackage{color}

\title[Statistical analysis of fireballs: Seismic signature survey]{Statistical analysis of fireballs: Seismic signature survey}

\author[Neidhart et al.]{T. Neidhart$^1$, K. Miljkovi\'c$^1$, E.K. Sansom$^1$, H.A.R. Devillepoix$^1$, T. Kawamura$^2$, J.-L. Dimech$^3$, M.A. Wieczorek$^4$, P.A. Bland$^1$
\affil{$^1$School of Earth and Planetary Sciences, Space Science and Technology Centre, Curtin University, Perth, Australia}%
\affil{$^2$Institut de Physique du Globe de Paris, France}
\affil{$^3$Geoscience Australia, Canberra, Australia}
\affil{$^4$Observatoire de Cote d’Azur, Laboratoire Lagrange, Nice, France}
}%

\jid{PASA}
\doi{10.1017/pas.\the\year.xxx}
\jyear{\the\year}

\newcommand{\degree}{$^\circ$}
\usepackage{aas_macros}
\usepackage{hyperref} 
\hypersetup{colorlinks,citecolor=blue,linkcolor=blue,urlcolor=blue}

\hypersetup{draft}

\begin{document}

\begin{frontmatter}
\maketitle

\begin{abstract}
Fireballs are infrequently recorded by seismic sensors on the ground. If recorded, they are usually reported as one-off events. This study is the first seismic bulk analysis of the largest single fireball data set, observed by the Desert Fireball Network (DFN) in Australia in the period 2014--2019. The DFN typically observes fireballs from cm-m scale impactors. We identified 25 fireballs in seismic time series data recorded by the Australian National Seismograph Network (ANSN). This corresponds to 1.8\% of surveyed fireballs, at the kinetic energy range of 10$^6$ to 10$^{10}$ J. The peaks observed in the seismic time series data were consistent with calculated arrival times of the direct airwave or ground-coupled Rayleigh wave caused by shock waves by the fireball in the atmosphere (either due to fragmentation or the passage of the Mach cone). Our work suggests that identification of fireball events in the seismic time series data depends both on physical properties of a fireball (such as fireball energy and entry angle in the atmosphere) and the sensitivity of a seismic instrument. This work suggests that fireballs are likely detectable within 200 km direct air distance between a fireball and seismic station, for sensors used in the ANSN. If each DFN observatory had been accompanied by a seismic sensor of similar sensitivity, 50\% of surveyed fireballs could have been detected. These statistics justify the future consideration of expanding the DFN camera network into the seismic domain.
\end{abstract}

\end{frontmatter}

\section{INTRODUCTION }
\label{sec:intro}

\begin{figure}
\begin{center}
\includegraphics[width=\columnwidth]{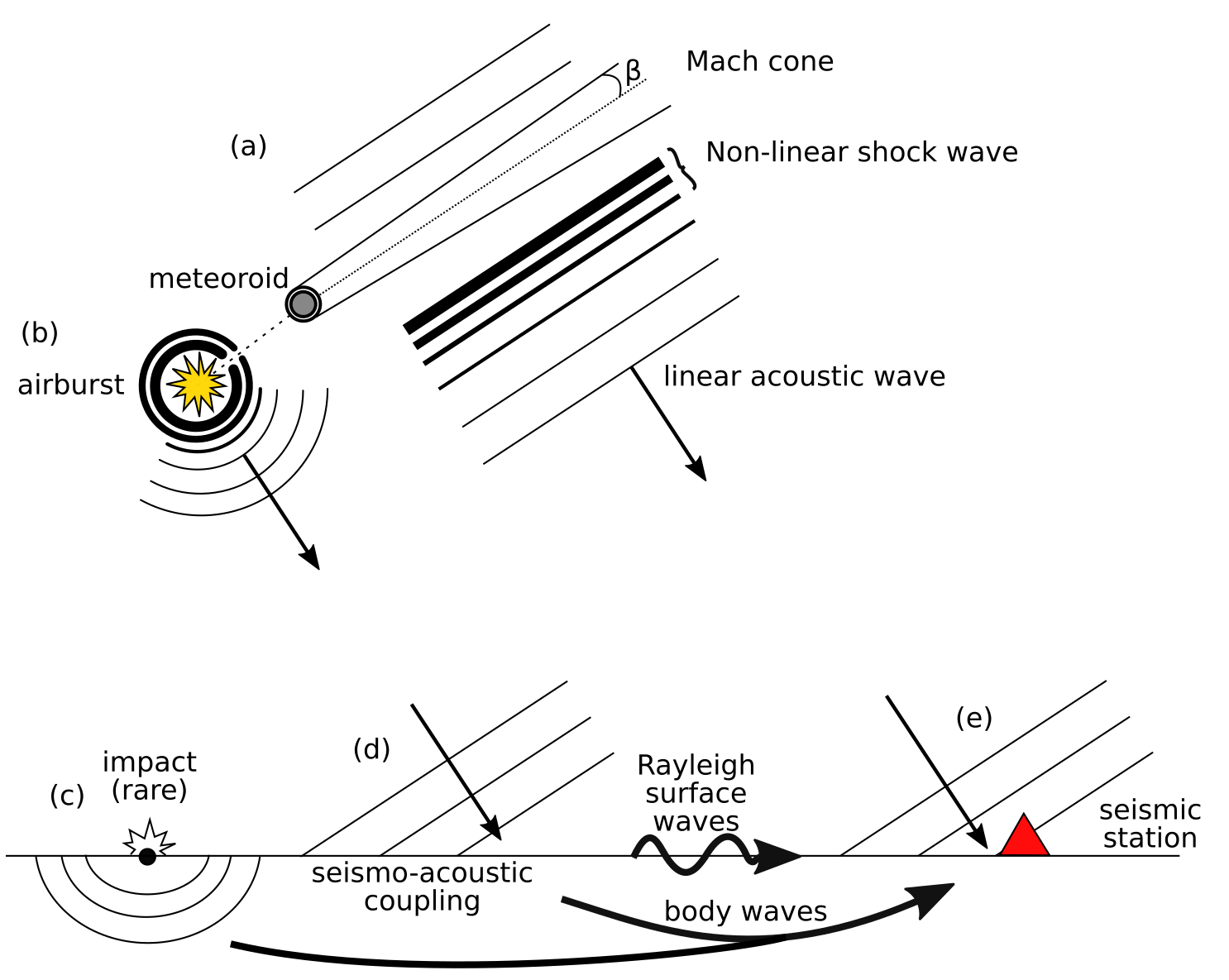}
\caption{Shock wave generation during a fireball event: (a) Shock waves are generated by the Mach cone that travel almost perpendicular to the trajectory of the object and rapidly decay from a non-linear to linear wavefront, (b) fragmentation-induced airburst causes shock waves that travel omnidirectionally, (c) seismic waves originating from impact itself (d) Rayleigh waves formed by coupling between airwaves and the ground, and (e) an air disturbance directed at the seismic station \citep{Brown2003,Revelle2004}. Figure redrawn from \cite{Edwards2008}.} \label{Edwards}
\end{center}
\end{figure}

When a meteoroid enters the atmosphere, it experiences aerodynamic drag and dynamic pressure. The atmosphere slows down meteoroids and in most cases they break-up and vaporize \citep{Ceplecha2005}. The break-up occurs when the dynamic pressure is higher than its compression strength \citep{Cevolani1994,Stevanovic2017}. Shock waves can be generated in the atmosphere by (Figure \ref{Edwards}):
\begin{itemize}
    \item The hypersonic flight forming a Mach cone,
    \item A discrete fragmentation event during the meteoroid's trajectory,
    \item A catastrophic final airburst,
    \item Physical impact on the ground (extremely rare).
\end{itemize} 
The Mach angle within the Mach cone is expected to be negligibly small, because the impact speed is much larger than the speed of sound in the air. Therefore, the shock waves generated during a hypersonic fireball entry are expected to propagate almost perpendicular to the trajectory (Figure \ref{Edwards}a). The fragmentation of the meteoroid can also create shock waves that propagate with no preferred direction; thus, can be assumed they propagate omnidirectionally (Figure \ref{Edwards}b). If the impactor or parts of the impactor survive the atmospheric path and hit the ground (Figure \ref{Edwards}c), the seismic waves in the ground can be generated by the impact itself \citep{Edwards2008,Tancredi2009}. The atmospheric shock waves can couple with the ground and form body and surface waves (Figure \ref{Edwards}d) \citep{Brown2003,Stevanovic2017,Karakostas2018}. The arrival times for different seismic waves differ as they travel at different speeds through different media (ground or air), which allows for their classification. Airwaves generated by the Mach cone (Figure \ref{Edwards}e) will arrive last as they travel slowest (at the speed of sound), through the air directly between the fireball and the sensor on the ground \citep{Edwards2008}.

For larger (bolide and cratering) events, a variety of seismic waves has been recorded. For example, the seismic signals caused by the 20-m diameter asteroid that exploded over Chelyabinsk, Russia in 2013 (estimated to have carried 10$^{15}$ J at airburst \citep{Emelyanenko2013}) were identified as P and S body waves, ground-coupled airwaves and Rayleigh waves \citep{Tauzin2013}; The P and S seismic waves were also seen when the 13.5-m diameter crater formed near Carancas, Peru in 2007 \citep{Brown2008, LePichon2008, Tancredi2009}; The Neuschwanstein large meteorite (estimated to have had 10$^{12}$ J initial source energy) \citep{Revelle2004, Oberst2004} caused seismic activity by direct airwaves and ground coupled Rayleigh waves at seismic stations within a few hundred km distance \citep{Revelle2004, Edwards2008}. These impact examples were all significantly larger than fireballs observed daily by the Desert Fireball Network (DFN) in Australia. Fireballs detected by the DFN have energies in the range of 10$^3$ to 10$^{12}$ J at atmospheric entry \citep{Devillepoix2019}. Meteorite-dropping fireballs are at the upper energy range observed by the DFN. 

DFN is the world’s largest fireball camera network, located in the Australian outback and consisting of 52 observatories, covering an area of 3 million km$^2$. It is aimed to detect fireballs, recover meteorites and to calculate their orbits \citep{Devillepoix2019,Devillepoix2018}. The observatories are optimised to image objects having a brightness between 0 to -15 magnitudes which corresponds to sizes between 0.05 and 0.5 m \citep{Devillepoix2019}. In this work, we make a bulk seismic analysis of the largest single data set of terrestrial fireballs obtained by the DFN in the period from 2014 to 2019, by systematically searching for seismic signals occurring in the time window and proximity of fireball trajectories. 

Unlike other studies that used data from images \citep{Beech1995,Brown1994,Spurny2012}, seismic stations \citep{Brown2003,Devillepoix2020,Koten2019} and infrasound \citep{ElGabry2017} to calculate the orbits and energies of meteors, this is the first study that uses information about the trajectory and timing of fireballs from a large dataset to back-trace any impact-related seismic activity. We investigate detection threshold of the DFN-observed fireballs in seismic data recorded by the Australian National Seismograph Network (ANSN). We also report on the seismic properties of the fireballs caught by the seismic instruments. This information will be used for future instrument development in detecting fireballs in the seismic domain. 

\section{Methodology}
 We used the DFN database containing trajectories of 1410 fireball events that occurred above Australia over the last 6 years. The DFN trajectory data provide absolute timing of fireball events, the start and end coordinates as well as the height above ground of the observed bright flight. A Python-based program was written to calculate distances between the entire fireball trajectory (bright flight path) and all ANSN seismic stations. The program was applied to all 1410 DFN fireballs. The arrival times for the airwave are then calculated for both the longest and the shortest direct distances, using a speed of sound of 300 $\pm$ 60 m/s. We used this error margin to account for local temperature and wind dependencies \citep{LePichon2008}. The large time window also considers unknown coupling with the ground and the low signal strength.
 
\begin{figure}
\centering
\includegraphics[width=\columnwidth]{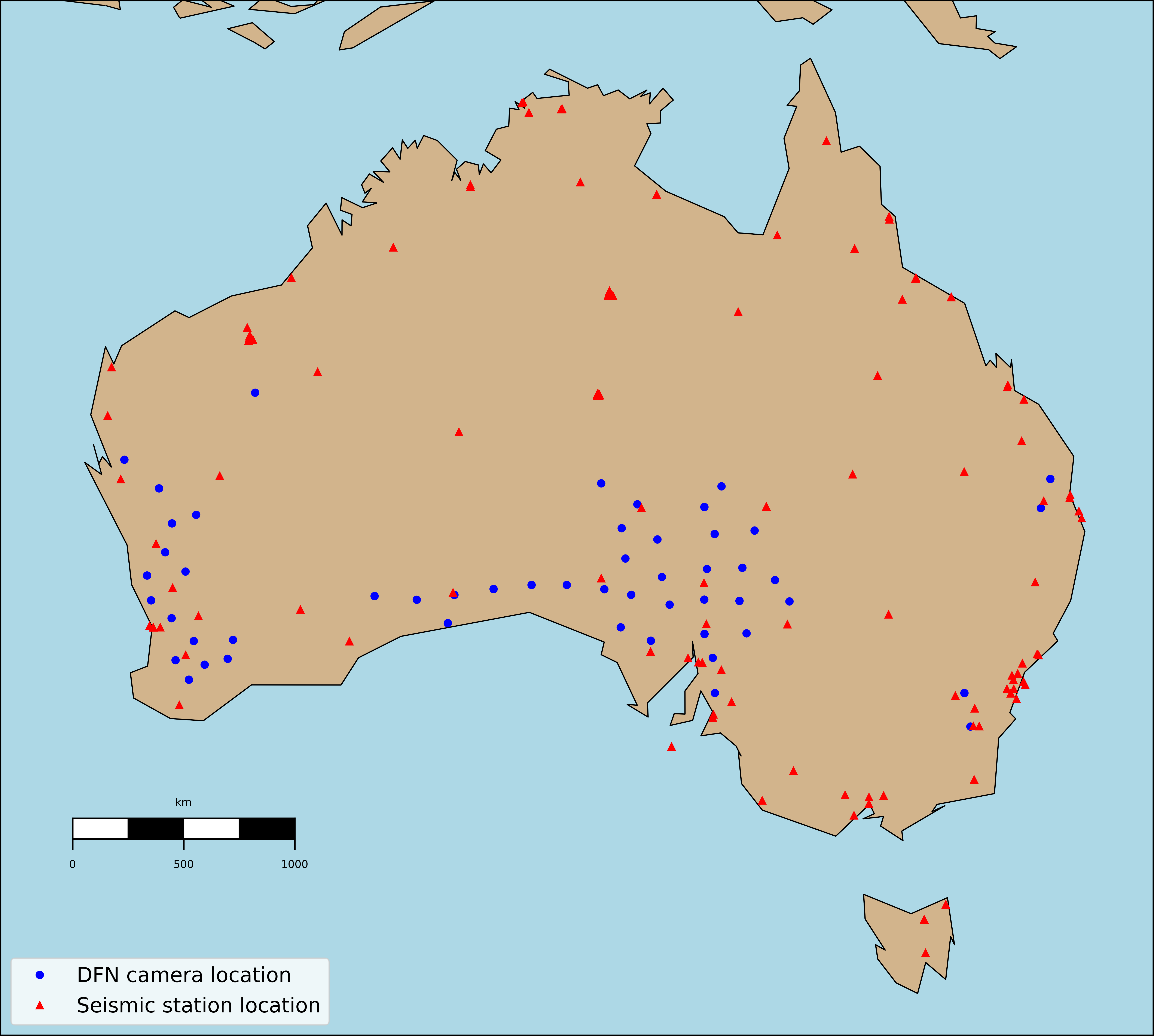}    
\caption{Locations of GA seismometers (red triangles) and DFN camera observatories (blue circles). Some stations are close together and therefore symbols overlap.}
\label{map}
\end{figure}

 Seismic data were acquired from the ANSN, operated by Geoscience Australia (GA), via public service domain IRIS (Incorporated Research Institutions for Seismology). The ANSN consists of a network of broadband seismometers across Australia and its offshore territories. Figure \ref{map} shows the locations of broadband seismometers (red triangles) and the coverage of DFN observatories (blue circles). 
 
The criteria that determined if a signal in time series data can be confidently classified as a signal coming from a fireball event are:
\begin{enumerate}
\item The amplitude of the signal must be similar or lower than previously confirmed seismic signals from fireballs or bolides, accounting for uncertainties related to the event’s distance to a detector, yet above the background noise;
\item The seismic signal must be within the calculated arrival times of the airwave (direct or ground-coupled Rayleigh wave; No P and S waves were identified in this survey);
\item There must not be an earthquake activity in the database (Geoscience Australia, 2019) at about the same time;
\item There must not be any clear anthropogenic-related noise (e.g., mine blasts, proximity to airport runways, etc). We note that DFN detects only night-time fireballs and at that time the anthropogenic noise is expected to be minimal.
\end{enumerate}

The seismic time series data were obtained from the nearest seismic stations and checked for distinguishable signals in the time window of the arrival of the airwave and Rayleigh wave (Criteria 2). Time series data were interrogated for a time window starting 30 seconds prior to the start of a fireball event in the upper atmosphere and ending up to 28 minutes later. This is to account for the travel time of the airwave from the fireball to any seismic station within 400 km. The seismic data was downloaded from the IRIS database. The Python framework ObsPy \citep{Beyreuther2010, Krischer2015} was used to manipulate and analyse the time series data and the Python library Astropy \citep{Astropy2013,Astropy2018} was used for making coordinate transformations. The time series data were filtered using a Butterworth-Highpass filter at a default frequency of 2 Hz. For most signals this filtering was the most satisfactory in cutting out ambient noise. 

In attempt to distinguish between meteor fragmentation and the Mach cone passage, we used two approaches. We looked into the fireball orientation with respect to the location of the seismic station. If the shortest distance to the seismic station is perpendicular to the bright flight trajectory and arrival time for the airwaves fits, signals are classified as likely originating from the Mach cone. If the shortest distance is not perpendicular to the bright flight trajectory, any seismic signals can be assumed to come from a fragmentation along the trajectory. Considering that the fragmentation has no preferred orientation, the events flagged as likely originated from the Mach cone could have instead originated from the airburst caused by fragmentation. However, we class them as Mach cone events because previous literature reported fragmentation to cause lesser air disturbance compared to the Mach cone passage \citep{Brown2003,Edwards2008}. We also visually investigated DFN fireball images to identify the distinct presence of fragmentation. However, we were unable to unambiguously make such a distinction for all fireball events. This is probably due to camera sensor saturation and because of DFN cameras using the deBruin shutter sequence to mark absolute timing which interrupts visual light curve recording (Table \ref{seismic_table}). 

\section{Results}

Compared to larger impact events, it was expected that the DFN-observed fireballs could only cause occasional weak seismic signals, predominantly coming from the atmospheric disturbance, and only in favourable positions and locations. Such an expectation was set by previous works \citep{Brown2003,Edwards2008}. 

Table \ref{DFNlist} shows the fireball events with suspected seismic signals including the start time of the bright flight observation. Seismic signals were found for 25 fireball events (Tables \ref{DFNlist}-\ref{seismic_table}) out of 1410 surveyed, setting the detectability at 1.8\% when using the public seismic data. From here on, we will refer to specific events with their allocated ID letter, rather than DFN event code name, as introduced in Table \ref{DFNlist}. 

\begin{figure}
\begin{center}
\includegraphics[width=\columnwidth]{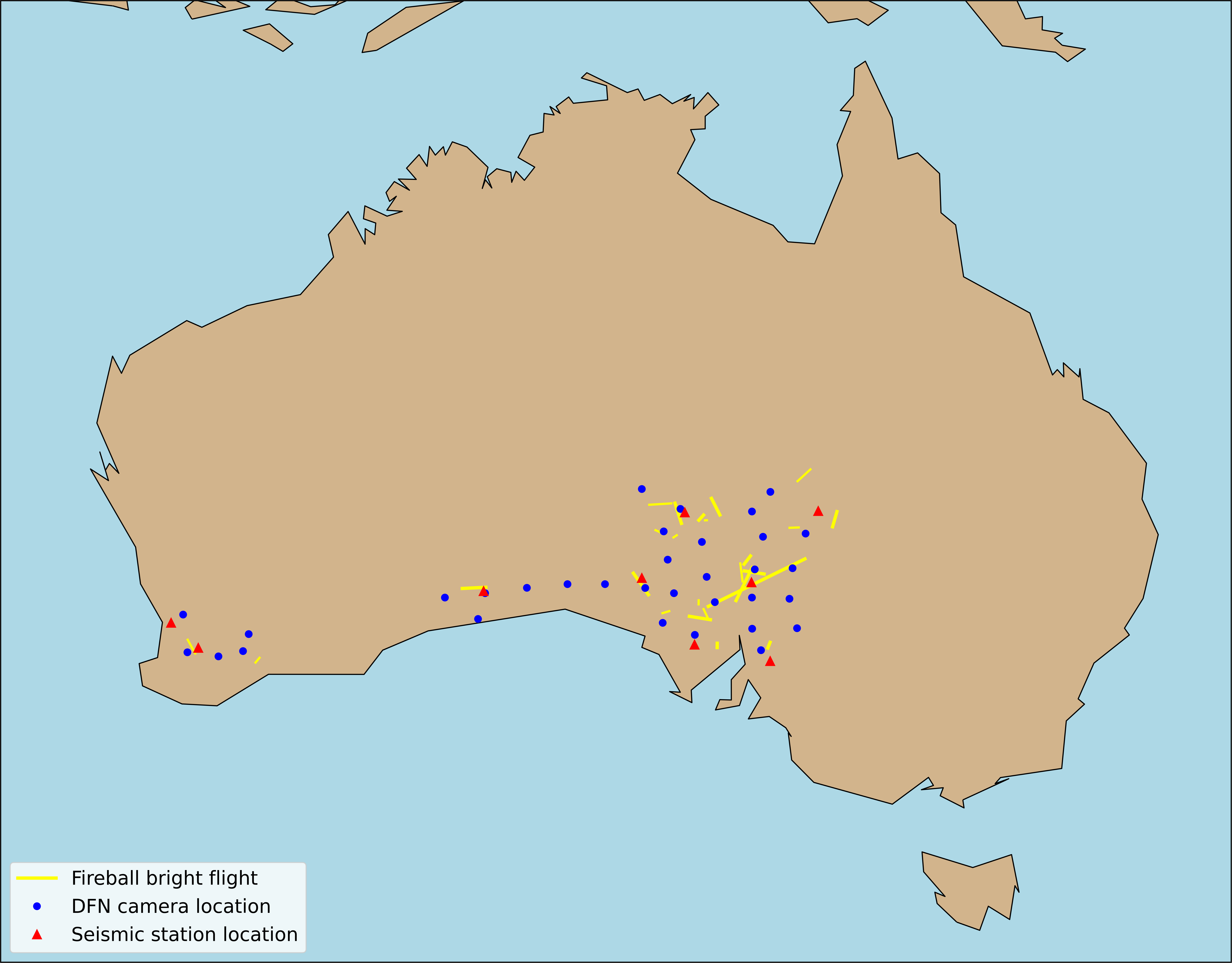}
\caption{Locations of seismic stations in Australia of the ANSN (red triangles) which detected seismic signals from fireballs, DFN observatories that observed fireballs that showed seismic signals (blue circles) and trajectories of the bright flight of fireballs for which suspected seismic signals have been detected (yellow lines). }\label{Fig_lines}
\end{center}
\end{figure}

Figure \ref{Fig_lines} shows the location of all DFN observatories (blue circles) and seismic stations of the ANSN (red triangles) that identified these 25 events. It also shows the trajectories of the bright flight of the fireballs for which seismic signals are suspected (yellow lines).

Table \ref{dfn_data} shows the coordinates of the beginning (lat$_b$, long$_b$) and the end (lat$_e$, long$_e$) of the bright flight, the beginning (h$_b$) and end (h$_e$) height, the trajectory slope, and the velocity (V), inferred mass (m) and fireball energy (KE) at atmospheric entry. The slope is defined as the angle between the beginning of the bright flight trajectory and local horizontal. The recorded fireballs had almost the entire range of possible impact angles (from 4\degree to 78\degree) with a mean value ($\pm1\sigma$) of 38\degree$\pm19$\degree. The mean h$_b$ was 86$\pm$25 km and h$_e$ was 46$\pm$18 km. The impact speed at the atmospheric entry was 25$\pm$13 km/s. Meteoroids had a very large mass range, from 1 g up to 180 kg estimated at atmospheric entry, corresponding to energies of 10$^6$ to 10$^{10}$ J. 

The peaks in the seismic time series data are consistent with the calculated arrival times of the airwave travelling perpendicular to the fireball trajectory and/or from an onmnidirectional source (fragmentation or frontal pressure at the end of the trajectory). Based on the orientation of the fireball trajectory with respect to the location of the nearest seismic station, 13 events [A:M] could have originated from the Mach cone shock wave (Figure \ref{Edwards}a) and 12 events [N:Y] were likely from an omnidirectional source (Figure \ref{Edwards}b; Tables 1-3).

\begin{table}
\caption{Fireball events with suspected seismic signals. Time of fireball marks the start of the bright flight as observed by the DFN. Notation [A:Y] is to be used for easier cross referencing between tables in this paper only.}
\centering
\begin{tabular}{lll}
\hline
\# & DFN event ID &      Time of fireball start \\
\hline
 A &  DN150622\_01 &  2015-06-22T10:55:08.126 \\
 B &  DN150822\_01 &  2015-08-22T11:42:14.446 \\
 C &  DN150829\_01 &  2015-08-29T18:11:09.126 \\
 D &  DN160210\_01 &  2016-02-10T11:48:49.126 \\
 E &  DN160328\_01 &  2016-03-28T12:19:02.286 \\
 F &  DN160604\_04 &  2016-06-04T20:17:01.126 \\
 G &  DN160610\_01 &  2016-06-10T16:19:22.726 \\
 H &  DN160802\_03 &  2016-08-02T13:24:51.926 \\
 I &  DN160918\_02 &  2016-09-18T10:48:37.426 \\
 J &  DN161007\_01 &  2016-10-07T19:01:01.026 \\
 K &  DN170723\_02 &  2017-07-23T13:16:47.800 \\
 L &  DN171214\_02 &  2017-12-14T15:29:42.026 \\
 M &  DN180421\_01 &  2018-04-21T15:03:42.800 \\
 N &  DN150909\_02 &  2015-09-09T12:08:31.126 \\
 O &  DN160620\_01 &  2016-06-20T11:25:37.426 \\
 P &  DN160830\_02 &  2016-08-30T09:35:20.926 \\
 Q &  DN160915\_01 &  2016-09-15T12:28:33.026 \\
 R &  DN161017\_01 &  2016-10-17T13:10:13.500 \\
 S &  DN161106\_03 &  2016-11-06T15:32:32.586 \\
 T &  DN161115\_03 &  2016-11-15T17:46:23.605 \\
 U &  DN170226\_03 &  2017-02-26T11:16:01.026 \\
 V &  DN170503\_04 &  2017-05-03T19:15:01.026 \\
 W &  DN170607\_01 &  2017-06-07T15:01:38.300 \\
 X &  DN171013\_03 &  2017-10-13T14:09:12.826 \\
 Y &  DN190407\_01 &  2019-04-07T13:20:32.150 \\
\hline
\end{tabular}
\label{DFNlist}
\end{table}

\begin{table*}
\caption{Fireball events with suspected seismic signals. Data includes the coordinates of the start (lat$_\text{b}$, long$_\text{b}$) and end (lat$_\text{e}$, long$_\text{e}$) of the bright flight trajectory, initial velocity (V), inferred mass (m), and fireball energies (KE) at the top of the atmosphere and slopes (with respect to the horizon) as observed by the DFN. The uncertainties in the trajectory positions are 0.1 km and the velocity uncertainties are 0.1 km/s. Masses are calculated using the dynamic method of \cite{Sansom2019} and are correct to an order of magnitude. Fireball energy is calculated as the transfer of kinetic energy on entry.}
\centering
\begin{tabular*}{\textwidth}{lllllrrrr}
\hline
ID & lat$_\text{b}$, long$_\text{b}$ & lat$_\text{e}$, long$_\text{e}$ &  h$_\text{b}$ &  h$_\text{e}$ &  V &  m &  KE &  Slope \\*
\, & (\degree,\degree) \, & (\degree,\degree) \, &  (km) &  (km) &   (km/s) &   (g) &   (MJ) &  (\degree)  \\
\hline
 A &                 -27.4556, 135.1753 &               -27.5068, 134.3404 &                                86.5 &                              29.8 &                     22.0 &      1200 &              286 &         34 \\
 B &                 -31.5395, 135.0993 &               -31.6170, 134.8425 &                                75.5 &                              42.9 &                     21.4 &      10 &                2.29 &         51 \\
 C &                 -30.4119, 137.8785 &               -29.7315, 137.7925 &                                69.6 &                              37.3 &                     15.3 &      900 &              105 &         23 \\
 D &                 -28.6623, 135.3836 &               -28.7460, 135.2587 &                                75.1 &                              41.7 &                     18.0 &      80 &               13 &         65 \\
 E &                 -30.6361, 128.1459 &               -30.6801, 127.2522 &                                99.2 &                              72.2 &                     38.9 &      2 &                1.51 &         17 \\
 F &                 -29.5592, 140.2430 &               -31.3590, 136.5810 &                                98.6 &                              66.8 &                     38.3 &      10 &                7.33 &          4 \\
 G &                 -31.8641, 136.6595 &               -31.7352, 135.8630 &                               111.0 &                              90.1 &                     69.9 &      1 &                2.44 &         15 \\
 H &                 -31.1390, 137.6323 &               -29.8901, 138.2426 &                                85.6 &                              31.4 &                     17.2 &      1500 &              222 &         19 \\
 I &                 -30.0159, 137.9459 &               -30.1322, 138.6963 &                                79.2 &                              41.4 &                     15.6 &      50 &                6.08 &         27 \\
 J &                 -28.5360, 134.7298 &               -28.4699, 134.5781 &                                60.4 &                              39.5 &                     38.3 &      4700 &             3450 &         51 \\
 K &                 -28.2104, 135.5600 &               -27.4458, 135.3177 &                                76.4 &                              30.0 &                     14.8 &      5500 &              602 &         28 \\
 L &                 -27.7673, 141.4574 &               -28.3419, 141.2917 &                                91.9 &                              54.9 &                     34.9 &      5 &                3.05 &         29 \\
 M &                 -30.1016, 133.7398 &               -30.9163, 134.3092 &                                88.2 &                              58.3 &                     14.4 &      30 &                3.11 &         16 \\
 N &                 -32.9211, 136.9160 &               -32.7568, 136.9180 &                                57.3 &                              45.0 &                     18.3 &      400 &               67 &         34 \\
 O &                 -28.3672, 140.0076 &               -28.3798, 139.6592 &                                97.2 &                              70.7 &                     31.5 &      60 &               29.80 &         37 \\
 P &                 -31.4627, 136.4002 &               -31.8320, 136.5779 &                                82.7 &                              42.0 &                     13.7 &      5800 &              544 &         42 \\
 Q &                 -26.1650, 140.4517 &               -26.6136, 139.9639 &                                90.4 &                              29.8 &                     29.3 &      900 &              408 &         41 \\
 R &                 -33.1445, 117.1267 &               -32.6246, 116.8404 &                                82.4 &                              25.6 &                     16.4 &      5300 &              713 &         42 \\
 S &                 -32.9373, 138.8363 &               -32.7216, 138.9215 &                               184.6 &                              63.0 &                     16.3 &      40 &                5.31 &         78 \\
 T &                 -27.8922, 137.0177 &               -27.2644, 136.6992 &                                78.1 &                              19.1 &                     13.1 &    180000 &            15400 &         37 \\
 U &                 -28.0908, 136.2213 &               -27.8967, 136.3981 &                                76.1 &                              56.6 &                     23.3 &      7 &                1.90 &         35 \\
 V &                 -28.0967, 136.4564 &               -28.0936, 136.5267 &                                54.3 &                              46.0 &                     34.7 &      1900 &             1140 &         50 \\
 W &                 -31.1292, 136.2143 &               -31.2810, 136.2115 &                                83.9 &                              27.6 &                     22.6 &      1500 &              383 &         73 \\
 X &                 -29.4472, 138.1771 &               -29.7606, 137.9361 &                                94.6 &                              54.5 &                     35.0 &      5 &                3.06 &         43 \\
 Y &                 -33.3076, 119.5644 &               -33.4876, 119.4131 &                                76.1 &                              29.7 &                     16.4 &      877 &              207.43 &         62 \\
\hline
\end{tabular*}
\label{dfn_data}
\end{table*}

\begin{table*}
\caption{Fireball events with suspected seismic signal data, including the shortest station-to-trajectory distance (d$_{\text{min}}$), peak values for the seismic acceleration in vertical (BHZ), N-S (BHN) and E-W (BHE) seismic axes, estimated duration of the seismic signal (t), and peak frequency ($\nu$) after applying 2 Hz high pass filter. Based on the arrival times, the seismic source can be a direct airwave (A) or a ground-coupled Rayleigh wave (R). The last column shows whether the optical image of the fireball displayed clear evidence of fragmentation processes. *Note that NWAO station is non-aligned to cardinals.}
\centering
\begin{tabular*}{\textwidth}{lllllrrrrr}
\hline
ID & Station & d$_\text{{min}}$ & BHZ $\times$ 10$^{-3}$ & BHN $\times$ 10$^{-3}$ & BHE $\times$ 10$^{-3}$ & t & $\nu$ & Source & Fragmentation \\*
\, & \, & (km) & (mm/s$^2$) &  (mm/s$^2$) &  (mm/s$^2$) &  (s) & (Hz) & \, \\
\hline
 A &             OOD &                   106.8 &           1.38 &      0.62 &      0.67 &            16 &                             3 &     A and/or R & yes \\
 B &            BBOO &                   180.7 &           3.91 &      6.54 &      1.78 &            12 &                          4-10 &              A & no \\
 C &            LCRK &                    74.2 &           9.79 &      5.71 &      6.56 &             7 &                             3 &         A or R & no \\
 D &             OOD &                   121.3 &           3.47 &      1.03 &      1.17 &             3 &                             3 &              A & yes \\
 E &            FORT &                    93.0 &           5.04 &      0.67 &      0.17 &             7 &                             3 &     A and/or R & yes \\
 F &            LCRK &                    78.9 &           3.21 &      2.22 &      2.88 &            20 &                             3 &         A or R & - \\
 G &            BBOO &                   150.3 &           1.01 &      0.26 &      0.66 &             9 &                           3-5 &         A or R & no \\
 H &            LCRK &                    53.4 &           13.0 &      1.38 &      6.34 &            25 &                           3-6 &         A or R & yes \\
 I &            LCRK &                    69.4 &           9.22 &      3.52 &      4.64 &             9 &                           3-5 &         A or R & no \\
 J &             OOD &                   138.1 &           3.47 &      1.37 &      1.84 &             7 &                             3 &              A & yes \\
 K &             OOD &                    54.4 &           14.5 &      3.08 &      2.38 &            37 &                           3-5 &         A or R & no \\
 L &            INKA &                   100.3 &           17.4 &      4.26 &      6.56 &            55 &                             3 &         A or R & - \\
 M &            MULG &                    77.4 &           17.5 &      15.7 &      7.14 &            16 &                           3-4 &         A or R & yes \\
 N &            BBOO &                    92.7 &           0.71 &      0.36 &      0.35 &            24 &                           3-5 &              A & - \\
 O &            INKA &                   140.3 &           6.29 &      2.96 &      2.00 &            18 &                           3-4 &              A & yes \\
 P &            BBOO &                   126.5 &           3.56 &      1.82 &      0.92 &            17 &                           3-5 &     A and/or R & yes \\
 Q &            INKA &                   150.3 &           7.91 &      3.43 &      3.29 &             8 &                           3-5 &              A & yes \\
 R &             MUN &                    96.8 &           0.94 &      1.16 &      0.76 &            10 &                           3-5 &              A & - \\
 S &             HTT &                   101.1 &           3.08 &      1.69 &      3.22 &             7 &                           3-4 &         A or R & - \\
 T &             OOD &                   117.6 &           5.68 &      3.41 &      1.98 &            11 &                          3-10 &              A & - \\
 U &             OOD &                    90.9 &           3.31 &      2.53 &      1.64 &             8 &                           3-4 &         A or R & - \\
 V &             OOD &                    99.3 &           2.13 &      2.19 &      2.25 &             6 &                           3-4 &              A & yes \\
 W &            BBOO &                   172.7 &           0.62 &      0.82 &      1.03 &            12 &                           3-4 &              A & yes \\
 X &            LCRK &                    97.7 &           0.51 &      0.59 &      0.54 &            43 &                           3-5 &         A or R & yes \\
 Y &            NWAO* &                   214.5 &           1.47 &      -    &         - &            20 &                          7-10 &              A & no \\
\hline
\end{tabular*}
\label{seismic_table}
\end{table*}

\begin{figure}
\begin{center}
\includegraphics[width=\columnwidth]{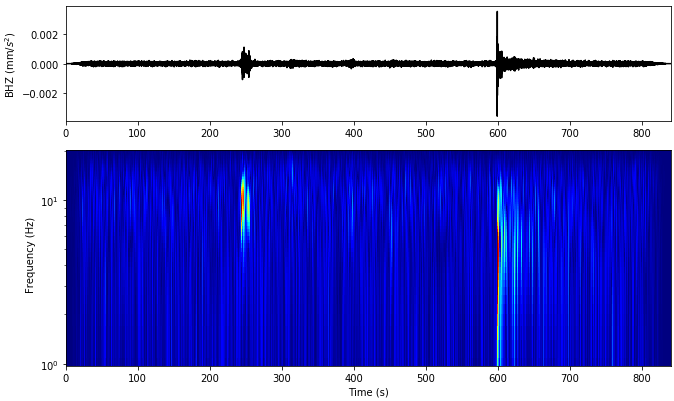}
\caption{Time series data and spectrogram in vertical direction for the only fireball event (DN160830\_02) for which signals of the airwave and the Rayleigh wave can be identified separately. Signal was detected at the stations BBOO and high pass filter was applied at 2 Hz.}\label{fig:spect}
\end{center}
\end{figure}

Figure \ref{fig:spect} shows one example of seismic time series data (top) and spectrogram (bottom) of fireball event P (Tables \ref{DFNlist} -- \ref{seismic_table}) for which signals of the airwave and the Rayleigh wave can be identified separately. Based on the seismic wave arrival time, the seismic source could either be from a direct airwave (A) or a ground-coupled Rayleigh wave (R). In some cases the arrival windows for A or R are clearly separated, but for most cases these windows overlap preventing us from confidently determining which source wave the signal came from (Table \ref{seismic_table}).

Table \ref{seismic_table} lists DFN fireball events for which we identified possible corresponding seismic signals, including the name of the seismic station at which the signal was detected, the shortest station-to-fireball distance (d$_{\text{min}}$), the peak values for the acceleration in vertical (BHZ), N-S (BHN) and E-W (BHE) components seen in the time series data, the duration of the signal (t), the peak frequency ($\nu$) and estimates for the seismic source. The seismic signals for all 25 fireballs are between 3 s and 55 s long and the peak values of the seismic frequencies are up to 10 Hz with an average at 3.8$\pm$1 Hz, which is in agreement with previous works \citep{Dauria2006,Edwards2008,Edwards2007, Kanamori1992,Revelle1976}. The shortest distance to the nearest seismic station is 112$\pm$40 km, ranging from 53 km to 215 km, although the surveyed area reached the maximum of 325 km distance. No surveyed fireballs were detected by more than one seismic station. This is expected given the sparse distribution of ANSN stations and is roughly in agreement with previous works \citep{Brown2003, Brown2004}.

Figure 5 shows the time series data for 25 fireball events [A:Y] for which seismic signals were detected. It can be seen that for 18 out of 25 events, the highest peaks are in the vertical direction. We examined any correlations between the direction of the highest peak in amplitude seen in the time series data and the position of the seismic station relative to the trajectory of the fireball and if the fireball approaches the seismic station or not. However, we did not find any other azimuth on directionality. On average, the amplitude for the highest peaks for seismic signals in the vertical direction was 5.5$\times$10$^{-3}$ mm/s$^2$, while it was 2.7$\times$10$^{-3}$ mm/s$^2$ in N-S and 2.4$\times$10$^{-3}$ mm/s$^2$ in E-W directions. This suggested a slight preference in vertical direction agreeing with the assumption that the seismic excitation was from the atmosphere. 

\begin{figure}
\centering
\begin{subfigure}{0.485\textwidth}
   \includegraphics[width=\linewidth]{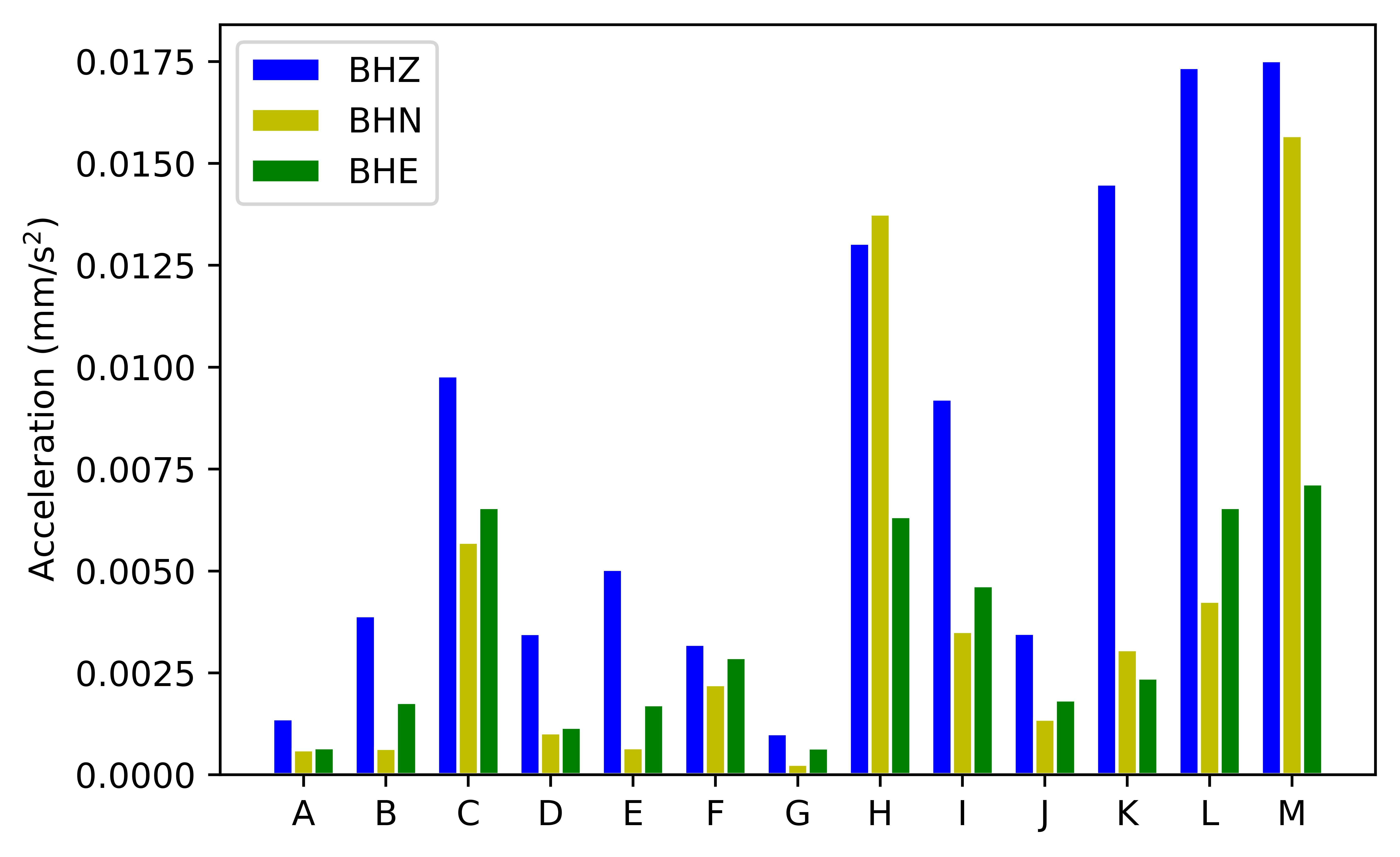}
   \label{Ng1} 
\end{subfigure}
\begin{subfigure}{0.485\textwidth}
   \includegraphics[width=\linewidth]{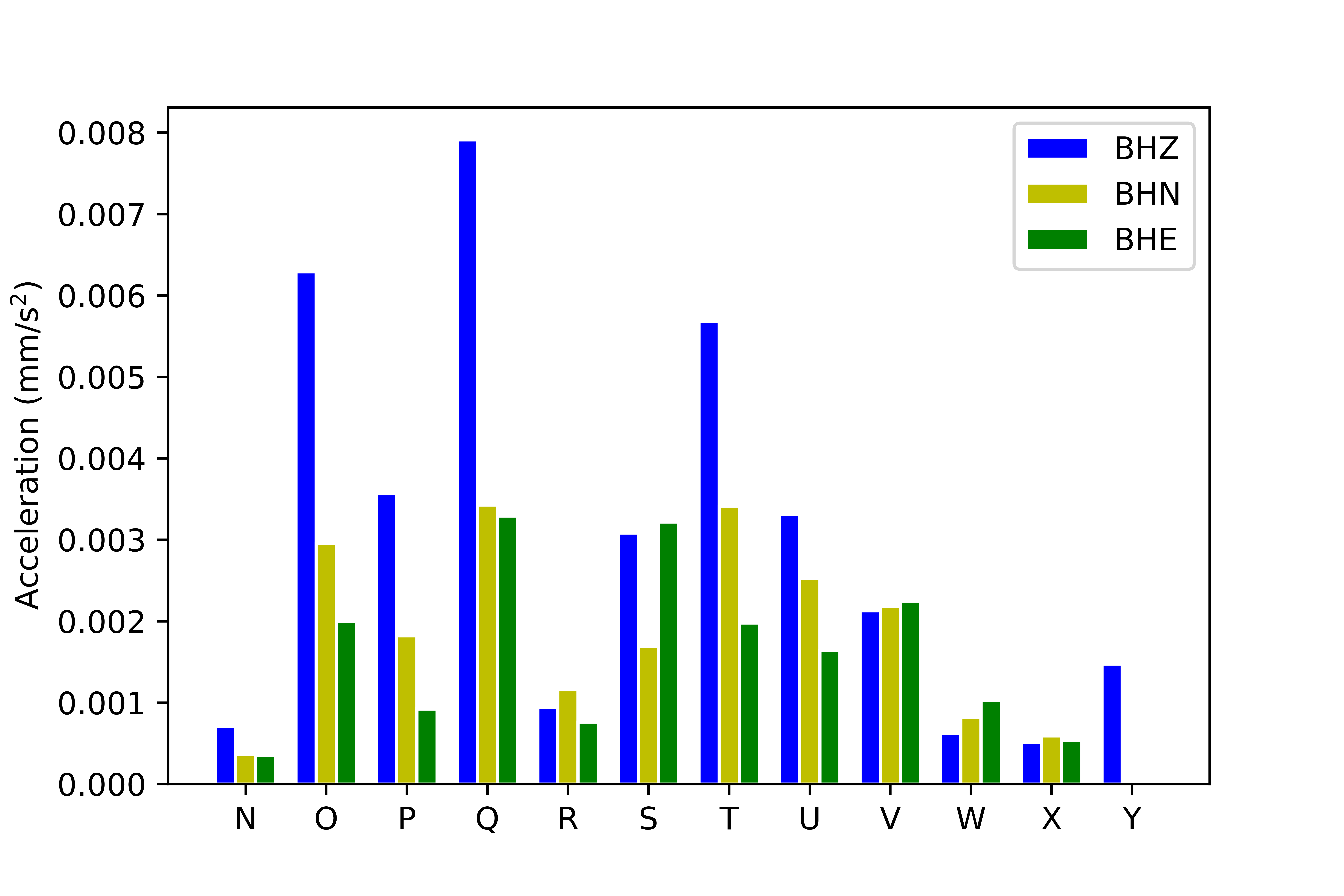}
   \label{Ng1}
\end{subfigure}

\caption{Highest peaks in time series data in vertical, North-South and East-West direction for 25 seismic signals that might originate from the Mach cone of fireball events (A-M) (upper) and from fragmentation (N-Y) (lower). 18 signals show the highest peak in vertical direction.}
\end{figure}

\begin{figure}
\begin{center}
\includegraphics[width=\columnwidth]{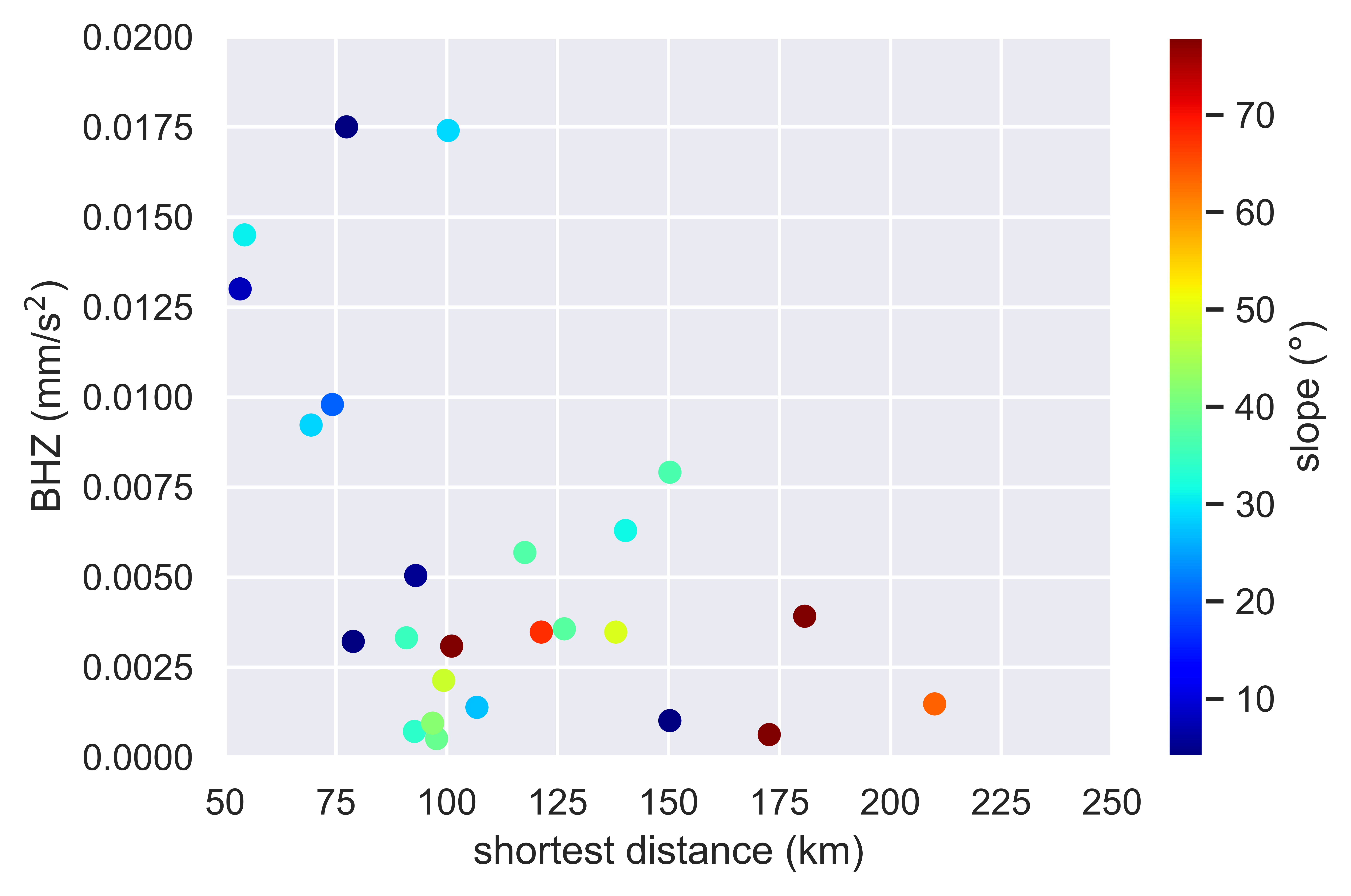}
\caption{Highest amplitude in vertical for all 25 fireballs for which seismic signals are suspected as a function of the shortest distance between bright flight trajectory and seismic station. The colours of the markers show the slope of these fireballs. The peak amplitude is decreasing with distance to the seismic station.}\label{Fig_correlation}
\end{center}
\end{figure}

Figure \ref{Fig_correlation} shows the highest peak in vertical direction as a function of the shortest distance between the trajectory and the seismic station for all events for which seismic signals are suspected. The colours of the markers represent the slope of the fireballs. It can be seen that fireballs that occur very close to the seismic station have higher peak amplitudes in vertical direction than fireballs further away. There is also additional observational bias that could be attributed to favourable fireball orientation to create Mach cone disturbance that is directed at a seismic station. The Mach cone-related fireball detections are more likely to originate from shallower (lower) impact angles that assure longer trajectories in the atmosphere than in the case of suspected fragmentation as a seismic source.

\section{Discussion} 

From the 1410 DFN fireball events surveyed, we identify seismic signals in time series data that correspond to 25 of these events. This is 1.8\%. Figure \ref{Fig_correlation} shows there is a rough correlation between peak amplitude and distance to a seismic station. Beyond 215 km we do not detect any unambiguous seismic signals, and the furthest events are all steep-sloped. It is therefore reasonable to place a threshold at 215 km as an approximate limit for the seismic detection of fireballs. Given that, the number of DFN fireballs within this range is reduced to 1101, increasing the detection success to 2.3\%. DFN observatories are approximately 150 km apart. There were 1236 of fireball trajectories within 215 km of a DFN observatory. Should the DFN camera network be equipped with seismic instruments (of comparable sensitivity) at each observatory site, 86\% of observed fireballs would be within the 215 km distance threshold for detection in the seismic domain. The mean distance to a seismic station of detected fireballs using ANSN was 112 km (Fig. \ref{Fig_correlation}), which corresponds to about 50\% of all surveyed fireballs if each observatory site had a seismic station equipped. It would be possible to detect fireballs at multiple stations, with an average of four stations per fireball. 

The survey showed that some seismic stations are more sensitive to fireball events than others. The highest number of signal detections was at the station Oodnadatta (OOD) which detected 7 suspected fireball events followed by Buckleboo (BBOO) and Leigh Creek (LCRK), where each detected 5 events, and Innamincka (INKA) with 3 events. There are five seismic stations (Forrest (FORT), Mundaring (MUN), Hallett (HTT), Mulgathing (MULG), Narrogin (NWAO)) that only detected one event. This could be due to the individual instrument quality or background noise levels which are influenced by the positioning setup and geographic location of the sensor. Previous studies by \cite{Revelle2004} have also pointed this out. Another sensitivity to detection might be directionality between seismic stations and bright flight trajectory. Seismic stations that are perpendicular can detect the signal from the Mach cone which has a higher amplitude and is therefore easier to recognize. A combination of these factors, like the presence of noise, distance to the station, the directionality from the trajectory to the seismic stations, weather conditions, soil properties and also the characteristics of the impactor, are among reasons we did not detect more than 2.3\% events within the 215 km threshold. 

As well as identifying the 25 fireball events in seismic time series data, we also investigated five of the largest events ever seen by the DFN. Unfortunately none pass the selection criteria. To date, there are two events detected by the DFN (DN150102\_01, DN170630\_01) that have also been recognized by the US Government Sensors (USG) and described in detail by \citet{Devillepoix2019}. The closest stations to these two events where data are available were 120 and 182 km away. These stations show noisy signals or a signal only in one component. 

We also looked for seismic signals from fireballs that had dropped a meteorite (Murrili, \cite{sansom2020murrili}; Dingle Dell, \cite{Devillepoix2018}; DN160822\_03, \cite{shober2019identification}) that were recovered from the field. The closest stations to these events were 150; 93 and 169; and 191 km respectively and show noisy seismic data and no signals. 

\section{Conclusions}

Fireball events occur on a daily basis, yet are rarely reported as seismic events because their energy (at the top of the atmosphere) is often not sufficient to cause quakes that are detectable by seismic stations. Unlike other studies who used data from images, seismic stations and infrasound to calculate the orbit and energies of meteors, this study uses information about the trajectory and timing of fireballs observed by the DFN to search for seismic signals. 

We report possible detections of 25 seismic signatures originating from 1410 surveyed fireballs observed by the DFN over a 6-year period. This is made by calculating the distance between the bright flight trajectory of the fireball to Australian National Seismograph Network (ANSN) seismic stations. We searched for significant seismic signals recorded that fit our selection criteria. The observed signals cannot be explained to be of any other geologic or anthropogenic origin. Signals are seconds-long in duration and have peak amplitude ranges in the following components:
\begin{itemize}
    \item Vertical: 5$\times10^{-4}$ mm/s$^2$ -- 2$\times10^{-2}$ mm/s$^2$
    \item N-S: 3$\times10^{-4}$ mm/s$^2$ -- 2$\times10^{-2}$ mm/s$^2$
    \item E-W: 4$\times10^{-4}$ mm/s$^2$ -- 7$\times10^{-3}$ mm/s$^{2}$
\end{itemize}

The total of 18 out of 25 signals showed the highest peak in vertical component. The signals showed the peak frequency in the range up to 10 Hz. Calculations of arrival times suggests signals are due to direct airwaves or ground-coupled Rayleigh waves. The fireball directionality suggest that about half of the observed signals could have been caused by the Mach cone and the other half originated from fragmentation of the impactor.  

We propose an upper threshold for seismic detectabilty of fireballs to be approximately 215 km. If a seismometer (of equal sensitivity) was installed alongside these systems, it may have been possible to record 50\% of all DFN fireballs.

\begin{acknowledgements}
TN is fully, and PAB partially, supported by the Australian Research Council on DP180100661. KM is fully supported by the Australian Research Council on DP180100661 and DE180100584. MW is supported by DP180100661 via Discovery International Award. The DFN, EKS, PAB and HARD would like to thank support from the Australian Research Council as part of the Australian Discovery Project scheme (DP170102529 and DP200102073), the institutional support from Curtin University, and the Pawsey Supercomputing Centre. This research made use of data from the Australian National Seismograph Network and are available through the IRIS Data Management Center (IRISDMC). The facilities of IRIS Data Services, and specifically the IRIS Data Management Center, were used for access to waveforms, related metadata, and/or derived products used in this study. IRIS Data Services are funded through the Seismological Facilities for the Advancement of Geoscience (SAGE) Award of the National Science Foundation under Cooperative Support Agreement EAR-1851048. This is InSight Contribution Number 168.
\end{acknowledgements}

\bibliographystyle{pasa-mnras}
\bibliography{1r_lamboo_notes}

\end{document}